\documentclass{llncs}

\usepackage{amssymb}
\usepackage{amsmath}
\usepackage{latexsym} 
\usepackage{epsfig}

\def\MYBOX{\medspace]\negthickspace\negmedspace[\medspace}
\def\Proof{\par\noindent{\bf Proof:}\indent}
\def\QED{\hfill$\Box$\par\vskip1em}

\begin{document}

\title{Discovering Network Topology in \\ 
       the Presence of Byzantine Faults\thanks{Some
       of the results in this article were presented
       at the 13th Colloquium on Structural Information and 
       Communication Complexity, Chester, UK in July 2006}}

\author{Mikhail Nesterenko\inst{1}\thanks{This author was supported 
in part by DARPA contract OSU-RF\#F33615-01-C-1901 and by NSF
CAREER Award 0347485. Part of this work was done while the author was visiting Paris Sud University.} 
\and
S\'ebastien Tixeuil\inst{2}\thanks{This author was supported in part by the 
FNS grants FRAGILE and SR2I from ACI ``S\'{e}curit\'{e} et Informatique'',
and ANR grant SOGEA from ARA program ``S\'{e}curit\'{e}, Syst\`{e}mes Embarqu\'{e}s et Intelligence Ambiante''. Part of this work was done while the author was visiting Kent State University. 
}
}
\institute{
Computer Science Department, 
Kent State University, 
Kent, OH, 44242, USA, 
\email{mikhail@cs.kent.edu}
\and
LRI-CNRS UMR 8623 \& INRIA Grand Large\\
Universit\'{e} Paris Sud, France, 
\email{tixeuil@lri.fr}
}

\maketitle

\begin{abstract}
We study the problem of Byzantine-robust topology discovery in an
arbitrary asynchronous network. We formally state the weak and strong
versions of the problem. The weak version requires that either each
node discovers the topology of the network or at least one node
detects the presence of a faulty node. The strong version requires
that each node discovers the topology regardless of faults.

\ \ \ \ We focus on non-cryptographic solutions to these problems. We
explore their bounds. We prove that the weak topology discovery
problem is solvable only if the connectivity of the network exceeds
the number of faults in the system. Similarly, we show that the strong
version of the problem is solvable only if the network connectivity is
more than twice the number of faults.

\ \ \ \ We present solutions to both versions of the problem. The
presented algorithms match the established graph connectivity
bounds. The algorithms do not require the individual nodes to know
either the diameter or the size of the network. The message complexity
of both programs is low polynomial with respect to the network
size. We describe how our solutions can be extended to add the
property of termination, handle topology changes and perform
neighborhood discovery.
\end{abstract}

\section{Introduction}
In this paper, we investigate the problem of Byzantine-tolerant
distributed topology discovery in an arbitrary network. Each node is
only aware of its neighboring peers and it needs to learn the topology
of the entire network.

Topology discovery is an essential problem in distributed computing
(\emph{e.g.} see~\cite{SG89}). It has direct applicability in
practical systems.  For example, link-state based routing protocols
such as OSPF use topology discovery mechanisms to compute the routing
tables. Recently, the problem has come to the fore with the
introduction of ad hoc wireless sensor networks, such as Berkeley
motes~\cite{Hill:2002:MWP}, where topology discovery is indispensable for
routing decisions.

As reliability demands on distributed systems increase, the interest
in developing robust topology discovery programs grows.  One of the
strongest fault models is \emph{Byzantine} \cite{LSP82}: the faulty
node behaves arbitrarily. This model encompasses rich set of fault
scenarios. Moreover, Byzantine fault tolerance has security
implications, as the behavior of an intruder can be modeled as
Byzantine.  One approach to deal with Byzantine faults is by enabling
the nodes to use cryptographic operations such as digital signatures
or certificates.  This limits the power of a Byzantine node as a
non-faulty node can verify the validity of received topology
information and authenticate the sender across multiple hops.
However, this option may not be available. For example, wireless
sensors may not have the capacity to manipulate digital signatures.
Another way to limit the power of a Byzantine process is to assume
synchrony: all processes proceed in lock-step.  Indeed, if a process
is required to send a message with each pulse, a Byzantine process
cannot refuse to send a message without being detected. However, the
synchrony assumption may be too restrictive for practical systems.

\ \\ \textbf{Our contribution.} In this study we explore the
fundamental properties of topology discovery. We select the weakest
practical programming model, establish the limits on the solutions and
present the programs matching those limits.

Specifically, we consider arbitrary networks of arbitrary topology
where up to fixed number of nodes $k$ is faulty. The execution model
is asynchronous. We are interested in solutions that do not use
cryptographic primitives. The solutions should be terminating and the
individual processes should not be aware of the network parameters
such as network diameter or its total number of nodes.

We state two variants of the topology discovery problem: \emph{weak}
and \emph{strong}. In the former --- either each non-faulty node
learns the topology of the network or one of them detects a fault; in
the latter --- each non-faulty node has to learn the topology of the
network regardless of the presence of faults. 

As negative results we show that any solution to the weak topology
discovery problem can not ascertain the presence of an edge between
two faulty nodes.  Similarly, any solution to the strong variant can
not determine the presence of a edge between a pair of nodes at least
one of which is faulty. Moreover, the solution to the weak variant
requires the network to be at least $(k+1)$-connected. In case of the
strong variant the network must be at least $(2k+1)$-connected.

The main contribution of this study are the algorithms that solve the
two problems: \emph{Detector} and \emph{Explorer}. The algorithms
match the respective connectivity lower bounds.  To the best of our
knowledge, these are the first asynchronous Byzantine-robust solutions
to the topology discovery problem that do not use cryptographic
operations.  \emph{Explorer} solves the stronger problem. However,
\emph{Detector} has better message complexity.  \emph{Detector} either
determines topology or signals fault in $O(\delta n^3)$ messages where
$\delta$ and $n$ are the maximum neighborhood size and the number of
nodes in the system respectively.  \emph{Explorer} finishes in
$O(n^4)$ messages.  We extend our algorithms to (a) terminate (b)
handle topology changes (c) discover neighbors if ports are known (d)
discover a fixed number of routes instead of complete topology and (e)
reliably propagate arbitrary information instead of topological 
data.

\ \\ \textbf{Related work.}  A number of researchers employ
cryptographic operations to counter Byzantine faults. Avromopolus et
al \cite{AKWK04} consider the problem of secure routing. Therein see
the references to other secure routing solutions that rely on
cryptography. Perrig et al \cite{Perrig:2004:SWS} survey robust
routing methods in ad hoc sensor networks. The techniques covered
there also assume that the processes are capable of cryptographic
operations.

A naive approach of solving the topology discovery problem without
cryptography would be to use a Byzantine-resilient broadcast
\cite{BV05,D82,K04,PP05}: each node advertises its neighborhood.
However all existing solutions for arbitrary topology known to us
require that the graph topology is \emph{a priori} known to the nodes.

Let us survey the non-cryptography based approaches to Byzantine
fault-tolerance.  Most programs described in the literature
\cite{AW98,MRRS01,MMR03,MS03} assume completely connected networks and
can not be easily extended to deal with arbitrary topology. Dolev
\cite{D82} considers Byzantine agreement on arbitrary graphs.  He
states that for agreement in the presence of up to $k$ Byzantine
nodes, it is necessary and sufficient that the network is
$(2k+1)$-connected and the number of nodes in the system is at least
$3k+1$.  However, his solution requires that the nodes are aware of
the topology in advance.  Also, this solution assumes the synchronous
execution model.  Recently, the problem of Byzantine-robust reliable
broadcast has attracted attention \cite{BV05,K04,PP05}. However, in
all cases the topology is assumed to be known. Bhandari and Vaidya
\cite{BV05} and Koo \cite{K04} assume two-dimensional grid. Pelc and
Peleg \cite{PP05} consider arbitrary topology but assume that each
node knows the exact topology a priori.  A notable class of algorithms
tolerates Byzantine faults locally \cite{MT05r,NA02,SOM04}. Yet, the
emphasis of these algorithms is on containing the fault as close to
its source as possible. This is only applicable to the problems where
the information from remote nodes is unimportant such as vertex
coloring, link coloring or dining philosophers. Thus, local
containment approach is not applicable to topology discovery.

Masuzawa \cite{M95} considers the problem of topology discovery and
update. However, Masuzawa is interested in designing a
self-stabilizing solution to the problem and thus his fault model is
not as general as Byzantine: he considers only transient and crash
faults.

\ \\ The rest of the paper is organized as follows.  After stating our
programming model and notation in Section~\ref{SecNotation}, we
formulate the topology discovery problems, as well as state the
impossibility results in Section~\ref{SecBounds}. We present
\emph{Detector} and \emph{Explorer} in Sections~\ref{SecDetector} and
\ref{SecExplorer} respectively.  We discuss the composition of our
programs and their extensions in Section~\ref{SecExtensions} and
conclude the paper in Section~\ref{SecEnd}.

\section{Notation, Definitions and Assumptions}
\label{SecNotation}

\noindent
\textbf{Graphs.} A distributed \emph{system} (or \emph{program})
consists of a set of processes and a \emph{neighbor} relation between
them. This relation is the system \emph{topology}. The topology forms
a graph $G$.  Denote $n$ and $e$ to be the number of nodes\footnote{We
use terms \emph{process} and \emph{node} interchangeably.} and edges
in $G$ respectively.  Two processes are \emph{neighbors} if there is
an edge in $G$ connecting them.  A set $P$ of neighbors of process $p$
is \emph{neighborhood} of $p$. In the sequel we use small letters to
denote singleton variables and capital letters to denote sets. In
particular, we use a small letter for a process and a matching capital
one for this process' neighborhood. Since the topology is symmetric,
if $q \in P$ then $p \in Q$. Denote $\delta$ to be the maximum number
of nodes in a neighborhood.

A \emph{node-cut} of a graph is the set of nodes $U$ such that $G
\setminus U$ is disconnected or trivial. A \emph{node-connectivity}
(or just \emph{connectivity}) of a graph is the minimum cardinality of
a node-cut of this graph. In this paper we make use of the following
fact about graph connectivity that follows from Menger's theorem (see
\cite{YG98}): if a graph is $k$-connected (where $k$ is some constant)
then for every two vertices $u$ and $v$ there exists at least $k$
internally node-disjoint paths connecting $u$ and $v$ in this graph.

\ \\ \textbf{Program model.}  A process contains a set of variables.
When it is clear from the context, we refer to a variable $var$ of
process $p$ as $var.p$. Every variable ranges over a fixed domain of
values.  For each variable, certain values are \emph{initial}.  Each
pair of neighbor processes share a pair of special variables called
\emph{channels}. We denote $Ch.b.c$ the channel from process $b$ to
process $c$. Process $b$ is the \emph{sender} and $c$ is the
\emph{receiver}. The value for a channel variable is chosen from the
domain of (potentially infinite) sequences of messages.

A \emph{state} of the program is the assignment of a value to every
variable of each process from its corresponding domain. A state is
\emph{initial} if every variable has initial value.  Each process
contains a set of actions. An action has the form $\langle name
\rangle : \langle guard \rangle \longrightarrow \langle command
\rangle$. A \emph{guard} is a boolean predicate over the variables of
the process.  A \emph{command} is sequence of assignment and branching
statements. A guard may be a receive-statement that accesses the
incoming channel. A command may contain a send-statement that modifies
the outgoing channel. A parameter is used to define a set of actions
as one parameterized action. For example, let $j$ be a parameter
ranging over values 2, 5 and 9; then a parameterized action $ac.j$
defines the set of actions $ac.(j=2)\ \MYBOX\ ac.(j=5)\ \MYBOX\
ac.(j=9)$.  Either guard or command can contain quantified constructs
\cite{dijkstra-scholten:1990a} of the form: $(\langle quantifier
\rangle \langle bound\ variables \rangle : \langle range \rangle :
\langle term \rangle)$, where \emph{range} and \emph{term} are boolean
constructs.

\ \\ \textbf{Semantics.}  An action of a process of the program is
\emph{enabled} in a certain state if its guard evaluates to
\textbf{true}.  An action containing receive-statement is enabled when
appropriate message is at the head of the incoming channel. The
execution of the command of an action updates variables of the
process.  The execution of an action containing receive-statement
removes the received message from the head of the incoming channel and
inserts the value the message contains into the specified
variables. The execution of send-statement appends the specified
message to the tail of the outgoing message.

A \emph{computation} of the program is a maximal fair sequence of
states of the program such that the first state $s_0$ is initial and
for each state $s_i$ the state $s_{i+1}$ is obtained by executing the
command of an action whose state is enabled in $s_i$. That is, we
assume that the action execution is \emph{atomic}.  The maximality of
a computation means that the computation is either infinite or it
terminates in a state where none of the actions are enabled. The
fairness means that if an action is enabled in all but finitely many
states of an infinite computation then this action is executed
infinitely often. That is, we assume \emph{weak fairness} of action
execution. Notice that we define the receive statement to appear as a
standalone guard of an action. This means, that if a message of the
appropriate type is at the head of the incoming channel, the receive
action is enabled. Due to weak fairness assumption, this leads to
\emph{fair message receipt} assumption: each message in the channel is
eventually received.  Observe that our definition of a computation
considers \emph{asynchronous} computations.

To reason about program behavior we define boolean predicates on
program states. A program \emph{invariant} is a predicate that is
\textbf{true} in every initial state of the program and if the
predicate holds before the execution of the program action, it also
holds afterwards. Notice that by this definition a program invariant
holds in each state of every program computation.

\ \\ \textbf{Faults.} Throughout a computation, a process may be
either Byzantine (faulty) or non-faulty. A Byzantine process contains
an action that assigns to each local variable an arbitrary value from
its domain. This action is always enabled. Yet, the weak fairness
assumption does not apply to this action. That is, we consider
computations where a faulty process does not execute any actions.
Observe that we allow a faulty node to send arbitrary messages. We
assume, however, that messages sent by such a node conform to the
format specified by the algorithm: each message carries the specified
number of values, and the values are drawn from appropriate domains.
This assumption is not difficult to implement as message syntax
checking logic can be incorporated in receive-action of each
process. We assume \emph{oral record} \cite{LSP82} of message
transmission: the receiver can always correctly identify the message
sender. The channels are reliable: the messages are delivered in FIFO
order and without loss or corruption.  Throughout the paper we assume
that the maximum number of faulty nodes in the system is bounded by
some constant $k$.

\ \\ \textbf{Graph exploration.} The processes discover the topology
of the system by exchanging messages. Each message contains the
identifier of the process and its neighborhood.  Process $p$
\emph{explored} process $q$ if $p$ received a message with $(q,Q)$.
When it is clear from the context, we omit the mention of $p$.  An
\emph{explored} subgraph of a graph contains only explored
processes. A Byzantine process may potentially circulate information
about the processes that do not exist in the system altogether.  A
process is \emph{fake} if it does not exist in the system, a process
is \emph{real} otherwise.

\section{The Topology Discovery Problem: Statement and Solution Bounds}
\label{SecBounds}

\textbf{Problem statement.}

\begin{definition}[Weak Topology Discovery Problem]\em
A program is a solution to the weak topology discovery problem if each
of the program's computation satisfies the following properties:
\emph{termination} --- either all non-faulty processes determine the
system topology or at least one process detects a fault; \emph{safety}
--- for each non-faulty process, the determined topology is a subset
of the actual system topology; \emph{validity} --- the fault is
detected only if there are faulty processes in the system.
\end{definition}

\begin{definition}[Strong Topology Discovery Problem]\em
A program is a solution to the strong topology discovery problem if
each of the program's computations satisfies the following properties:
\emph{termination} --- all non-faulty processes determine the system
topology; \emph{safety} --- the determined topology is a subset of the
actual system topology.
\end{definition}

According to the safety property of both problem definitions each
non-faulty process is only required to discover a subset of the actual
system topology. However, the desired objective is for each node to
discover as much of it as possible. The following definitions capture
this idea.  A solution to a topology discovery problem is
\emph{complete} if every non-faulty process always discovers the
complete topology of the system. A solution to the problem is
\emph{node-complete} if every non-faulty process discovers all nodes
of the system. A solution is \emph{adjacent-edge complete} if every
non-faulty node discovers each edge adjacent to at least one
non-faulty node. A solution is \emph{two-adjacent-edge complete} if
every non-faulty node discovers each edge adjacent to two non-faulty
nodes.

\ \\ \textbf{Solution bounds.}  To simplify the presentation of the
negative results in this section we assume more restrictive execution
semantics. Each channel contains at most one message. The computation
is synchronous and proceeds in rounds.  In a single round, each
process consumes all messages in its incoming channels and outputs its
own messages into the outgoing channels.  Notice that the negative
results established for this semantics apply for the more general
semantics used in the rest of the paper.

\begin{theorem}\label{NoWeakComplete}\em
There does not exist a complete solution to the weak topology 
discovery problem.
\end{theorem}

\Proof Assume there exists a complete solution to the problem.
Consider $k \geq 2$ and topology $G_1$ that is not completely
connected. Let none of the nodes in $G_1$ be faulty. By the validity
property, none of the nodes may detect a fault in such
topology. Consider a computation $s_1$ of the solution program where
each node discovers $G_1$.  Let $p \in G_1$, $q \neq p$, and $r \neq
p$ be three nodes in $G_1$, with $q$ and $r$ being non-neighbor nodes
in $G_1$.  Since $G_1$ is not completely connected we can always find
two such nodes.

We form topology $G_2$ by connecting $q$ and $r$ in $G_1$.  Let $q$
and $r$ be faulty in $G_2$. We construct a computation $s_2$ which is
identical to $s_1$. That is, $q$ and $r$, being faulty, in every round
output the same messages as in $s_1$. Since $s_2$ is otherwise
identical to $s_1$, process $p$ determines that the topology of the
system is $G_1 \neq G_2$. Thus, the assumed solution is not
complete. \QED

\begin{theorem} \label{NoWeak}\em
There exists no node- and adjacent-edge complete solution to the weak
topology problem if the connectivity of the graph is lower or equal to
the total number of faults $k$.
\end{theorem}

\Proof
Assume the opposite. Let there be a node- and adjacent-edge complete
program that solves the problem for graphs whose connectivity is $k$
or less. Let $G_1$ and $G_2$ be two graphs of connectivity $k$.

This means that $G_1$ and $G_2$ contain the respective cut node sets
$A_1$ and $A_2$ whose cardinality is $k$. Rename the processes in
$G_2$ such that $A_1=A_2$.  By definition $A_1$ separates $G_1$ into
two disconnected sets $B_1$ and $C_1$. Similarly, $A_2$ separates
$G_2$ into $B_2$ and $C_2$.  Assume that $B_1 \not \subseteq B_2$. Since
$A_1=A_2$ we can form graph $G_3$ as $A_1 \cup B_2 \cup C_1$.

Let $s_1$ be any computation of the assumed program in the system of
topology $G_1$ and no faulty nodes. Since the program solves the weak
topology problem, the computation has to comply with all the
properties of the problem. By validity property, no fault is detected
in $s_1$. By termination property, each node in $G_1$, including
some node $p \in C_1$, eventually discovers the system topology.

By safety property the topology discovered by $p$ is a subset of
$G_1$.  Since the solution is complete the discovered topology is
$G_1$ exactly. Let $s_2$ be any computation of the assumed program in
the system of topology $G_2$ and no faulty nodes. Again, none of the
nodes detects a fault and all of them discover the complete topology
of $G_2$ in $s_2$.

We construct a new computation $s_3$ of the assumed program as
follows.  The system topology for $s_3$ is $G_3$ where all nodes in
$A_1$ are faulty.  Each faulty node $q \in A_1$ behaves as follows.
In the channels connecting $q$ to the nodes of $C_1 \subset G_3$, each
round $q$ outputs the messages as in $s_1$. Similarly, in the channels
connecting $q$ to the nodes of $B_2 \subset G_3$, $q$ outputs the
messages as in $s_2$. The non-faulty nodes of $B_2$ and $C_1$ behave
as in $s_1$ and $s_2$ respectively.

Observe that for the nodes of $B_2$, the topology and communication is
indistinguishable from that of $s_2$.  Similarly, for the nodes of
$C_1$ the topology and communication is indistinguishable from that of
$s_1$. Notice that this means that none of the non-faulty nodes detect
a fault in the system. Moreover, node $p \in C_1$ decides that the
system topology is the subset of $G_1$.  Yet, by construction, $G_1
\neq G_3$.  Specifically, $B_1 \not \subseteq B_2$. Moreover, none of
the nodes in $B_2$ are faulty.  If this is the case then either $s_3$
violates the safety property of the problem or the assumed solution is
not adjacent-edge complete.  The theorem follows.  \QED

Observe that for $(k+1)$-connected graphs an adjacent-edge complete solution
is also node complete.

\begin{theorem} \label{NoStrongEdgeComplete} \em
There does not exist an adjacent-edge complete solution to the strong
topology discovery problem.
\end{theorem}

\Proof Assume such a solution exists.  Consider system graph $G_1$
that is not completely connected. Let $p \in G_1$ be an arbitrary
node. Let $q \neq p$ and $r \neq p$ be two non-neighbor nodes of
$G_1$. We form topology $G_2$ by connecting $q$ and $r$ in $G_1$.

We construct computations $s_1$ and $s_2$ as follows. Let $s_1$ and
$s_2$ be executed on $G_1$ and $G_2$ respectively. And let $q$ be
faulty in $s_1$ and $r$ be faulty in $s_2$.  Set the output of $q$ in
each round to be identical in $s_1$ and $s_2$. Similarly, set the
output of $r$ to be identical in both computations as well. Since the
output of $q$ and $r$ in both computations is identical, we construct
the behavior of the rest of the nodes in $s_1$ and $s_2$ to be the
same.

Due to termination property, $p$ has to decide on the system topology
in both computations. 
Due to the safety property, in $s_1$ process $p$ has to determine that
the topology of the graph is a subset of $G_1$.  However, since the
behavior of $p$ in $s_2$ is identical to that in $s_1$, $p$ decides
that the topology of the system graph is $G_1$ in $s_2$ as well.  This
means $p$ does not include the edge between $q$ and $r$ to the
explored topology in $s_2$.  Yet, one of the nodes adjacent to this
edge, namely $q$, is not faulty. An adjacent-edge complete program
should include such edges in the discovered topology. Therefore, the
assumed program is not adjacent-edge complete.  \QED

\begin{theorem}\label{NoStrong}\em
There exists no node- and two-adjacent-edge complete solution to the
strong topology problem if the connectivity of the graph is less than
or equal to twice the total number of faults $k$.
\end{theorem}

\Proof Assume that there is a program that solves the problem for
graphs whose connectivity is $2k$ or less. Let $G_1$ and $G_2$ be two
different graphs whose connectivity is $2k$. Similar to the the proof
of Theorem~\ref{NoWeak}, we assume that $G_1 = A_1 \cup B_1 \cup C_1$
and $G_2 = A_2 \cup B_2 \cup C_2$ where the cardinality of $A_1$ and
$A_2$ are $2k$, $A_1 = A_2$, $B_1 \cap C_1 = \varnothing$, $B_2 \cap
C_2 = \varnothing$, and $B_1 \not \subseteq B_2$. Form $G_3 = A_1 \cup
B_2 \cup C_1$. Divide $A_1$ into two subsets $A'_1$ and $A''_1$ of the
same number of nodes.

Construct a computation $s_1$ with system topology $G_1$ where all
nodes in $A'_1$ are faulty; and another computation $s_3$ with system
topology $G_3$ where all nodes in $A''_1$ are faulty. The faulty nodes
in $s_1$ in the channels connecting $A'_1$ to $C_1$ communicate as the
(non-faulty) nodes of $A'_1$ in $s_3$. Similarly, the faulty nodes in
$s_3$ in the channels connecting $A''_1$ to $C_1$ communicate as the
nodes of $A''_1$ in $s_1$.  Observe that $s_1$ and $s_3$ are
indistinguishable to the nodes in $C_1$. Let the nodes in $C_1$,
including $p \in C_1$ behave identically in both
computations. According to the termination property of the strong
topology discovery problem every node, including $p$ has to determine
the system topology in both $s_1$ and $s_3$. Due to safety, the
topology that $p$ determines in $s_1$ is a subset of $G_1$. However,
$p$ behaves identically in $s_3$.

This means that $p$ decides that the system topology in $s_3$ is also
a subset of $G_1$. Since $G_1 \neq G_3$ (specifically, $B_1 \not
\subseteq B_2$), and that none of the nodes in $B_2$ are faulty, this
implies that either $s_3$ violates the safety property of the problem
or the assumed solution is not adjacent-edge complete.  The theorem
follows.
\QED

\section{Detector}
\label{SecDetector}

\noindent
\textbf{Outline.}  
\emph{Detector} solves the weak topology discovery problem for system
graphs whose connectivity exceeds the number of faulty nodes $k$.  The
algorithm leverages the connectivity of the graph.  For each pair of
nodes, the graph guarantees the presence of at least one path that
does not include a faulty node. The topology data travels along every
path of the graph.  Hence, the process that collects information about
another process can find the potential inconsistency between the
information that proceeds along the path containing faulty nodes and
the path containing only non-faulty ones.

Care is taken to detect the fake nodes whose information is introduced
by faulty processes. Since the processes do not know the size of the
system, a faulty process may potentially introduce an infinite number
of fake nodes. However, the graph connectivity assumption is used to
detect fake nodes. As faulty processes are the only source of
information about fake nodes, all the paths from the real nodes to the
fake ones have to contain a faulty node. Yet, the graph connectivity
is assumed to be greater than $k$. If a fake node is ever introduced,
one of the non-faulty processes eventually detects a graph with too
few paths leading to the fake node.

\begin{figure*}[ht]
\begin{tabbing}
1\=12345\=12345\=12345\=12345\=12345\=12345\=12345\=12345\=12345\=\kill

\> $\textbf{process}\ p$ \\ 
\> $\textbf{const}$ \\ 

\> \>  $P$: set of neighbor identifiers of $p$\\
\> \>  $k$: integer, upper bound on the number of faulty processes \\

\> $\textbf{parameter}$ \\
\> \>  $q: P$   \\

\> $\textbf{var}$ \\
\> \>$detect:$ boolean, initially \textbf{false}, signals fault \\
\> \>$start:$ boolean, initially \textbf{true}, 
      controls sending of  $p$'s neighborhood info\\
\> \> $TOP:$ set of tuples, initially $\{(p,P)\}$, 
  (process ids, neighbor id set) \\
\> \>\>\>\>\>\>     received by $p$\\
\> \>\>$*[$ \\
\>     \emph{init}:\>\>\>$start \longrightarrow $ \\
\> \>\>\>\>$start := \textbf{false}$, \\
\> \>\>\>\>$(\forall j: j \in P :\textbf{send}\ (p,P)\ \textbf{to}\ j)$\\
\> \>\>\ \ \MYBOX\\
\> \emph{accept}:\>\>\>$\textbf{receive}\ (r,R)\ \textbf{from}\ q \longrightarrow$ \\
\> \>\>\>\>$\textbf{if}\ (\exists s,S: (s,S) \in TOP: s=r 
                                                   \wedge S \neq R)\ \vee$ \\
\> \>\>\>\>\ \ \            $(\textbf{path\_number}(TOP \cup \{(r,R)\}) < k+1)$ \\
\> \>\>\>\>$\textbf{then}$ \\
\> \>\>\>\>\>$detect:=\textbf{true}$ \\
\> \>\>\>\>$\textbf{else}$ \\
\> \>\>\>\>\>$\textbf{if}\ (\nexists s,S: (s,S) \in TOP : s=r)\ 
                                                              \textbf{then}$ \\
\> \>\>\>\>\>\>$TOP := TOP \cup \{(r,R)\}$, \\
\> \>\>\>\>\>\>$(\forall j: j \in P :\textbf{send}\ (r,R)\ \textbf{to}\ j)$ \\
\> \>\>\ \ $]$ \\
\end{tabbing}
\caption{Process of \emph{Detector}} \label{FigDetector}
\end{figure*}

\ \\ \textbf{Detailed Description.}  The program is shown in
Figure~\ref{FigDetector}. Each process $p$ stores the identifiers of
its immediate neighbors. They are kept in set $P$. Each process keeps
the upper bound $k$ on the number of faulty processes.  Process $p$
maintains the following variables. Boolean variable $detect$ indicates
if $p$ discovers a fault in the system.  Boolean variable \emph{start}
guards the execution of the action that sends $p$'s neighborhood
information to its neighbors. Set $TOP$ (for topology) stores the
subgraph explored by $p$; $TOP$ contains tuples of the form:
(\emph{process identifier, its neighborhood}).  In the initial state,
$TOP$ contains $(p,P)$.

Function \textbf{path\_number} evaluates the topology of the subgraph
stored in $TOP$.  Recall that a node $u$ is unexplored by $p$ if for
every tuple $(s,S) \in TOP$, $s$ is not the same as $u$. That is $u$
may appear in $S$ only. We construct graph $G'$ by adding an edge to
every pair of unexplored processes present in $TOP$. We calculate the
value of \textbf{path\_number} as follows. If the information of $TOP$
is inconsistent, that is:
\[
\begin{split}
& (\exists u,v, U,V: ((u,U) \in TOP) \wedge ((v,V) \in TOP) :  \\
& (u \in V) \wedge (v \not\in U))
\end{split}
\]
\noindent
then \textbf{path\_number} returns $0$.  If there is exactly one
explored node in $TOP$, \textbf{path\_number} returns $k+1$. Otherwise
the function returns the minimum number of internally node disjoint
paths between two explored nodes in $G'$.  In the correctness proof
for this program we show that unless there is a fake node, the
\textbf{path\_number} of $G'$ is no smaller than the connectivity of
$G$.

Processes exchange messages of the form (\emph{process identifier, its
neighborhood id set}). A process contains two actions: \emph{init} and
\emph{accept}. Action \emph{init} starts the propagation of $p$'s
neighborhood throughout the system. Action \emph{accept} receives the
neighborhood data of some process, records it, checks against other
data already available for $p$ and possibly further disseminates the
data. If the data received from neighbor $q$ about a process $r$
contradicts what $p$ already holds about $r$ in $TOP$ or if the newly
arrived information implies that $G$ is less than $(k+1)$-connected
$p$ indicates that it detected a fault by setting $detect$ to
$\textbf{true}$.  Alternatively, if $p$ did not previously have the
information about $r$, $p$ updates $TOP$ and sends the received
information to all its neighbors.

 Observe that the propagation of
information about the neighborhood of a certain process is independent
of the information propagation of another process. Thus, we will focus
on the propagation of the information about a particular non-faulty
process $a$.

Let $COR$ contain each process $b$ such that $b$ is not faulty and
$TOP.b$ holds $(a,A)$. Let $a$ itself belong to $COR$ if $start.a$ is
\textbf{false}.

\begin{lemma} \label{ProveInvDetect}\em
The following predicate is an invariant of \emph{Detector}.
\begin{equation}\label{InvDetect}
\begin{split}
  &( \forall\ \textup{non-faulty}\  b,c: b \in COR, c \in B: \\
  &(c \in COR) \vee \\
  &((a,A) \in Ch.b.c))\ \vee \\
  &(\exists\ \textup{non-faulty}\ 
    j: j \in N: detect.j = \textbf{\textup{true}}) 
\end{split}
\end{equation}
\end{lemma}

The predicate states that unless one of the non-faulty processes in
the program detects a fault, if a process $b$ belongs to $COR$ then
each neighbor $c$ of $b$ either belongs to $COR$ as well or the
channel from $b$ to $c$ contains $(a,A)$.

\ \\
\Proof To prove that Predicate \ref{InvDetect} is an invariant of
the program, we need to show that it holds in the initial state of any
computation and it is closed under the execution of actions of
Byzantine as well as non-faulty processes.  The predicate holds
initially as the first disjunct is vacuously true. 

Note that no action of a Byzantine process immediately affects the
validity of the predicate.  Observe also that a non-faulty process can
only set $detect$ to $\textbf{true}$. Thus, once this happens the
predicate holds throughout the rest of the computation. Suppose
$detect$ is $\textbf{false}$ in all processes of the program. Then the
predicate is violated only if there is a non-faulty pair of neighbors
$b$ and $c$ such that $b$ belongs to $COR$, $c$ does not and there is
no message $(a,A)$ in the channel from $b$ to $c$. Notice that a
non-faulty process adds the first value $(r,R)$ to $TOP$ and never
changes it afterwards. Thus, provided that $detect=\textbf{false}$, to
violate the predicate, a process has to join $COR$ without sending
$(a,A)$ to its neighbors or consume a message with $(a,A)$ without
joining $COR$. Let us examine the actions of a non-faulty process and
ensure that neither of this happens.

Observe that \emph{init} is only of interest in $a$.  This action sets
$start.a = \textbf{false}$ which, by definition, adds $a$ to
$COR$. Also, \emph{init} atomically sends $(a,A)$ to all neighbors of
$a$.  Thus, the predicate is not violated by the execution of
\emph{init}.

Let us now consider \emph{accept} in an arbitrary non-faulty process
$u$. Let the message received by $u$ carry $(r,R)$. Observe that
\emph{accept} affects Predicate \ref{InvDetect} only if
$r=a$. \emph{accept} may make $u$ join $COR$ or consume a message with
$(a,A)$.  Notice, that if $u$ is already in $COR$ the receipt of a
message with $(a,A)$ does not violate the predicate. Also, $u$ joins
$COR$ only if it receives $(a,A)$. Hence, the only case we have to
consider is when $u$ does not belong to $COR$ before the execution of
\emph{accept}, $u$ receives $(a,A)$ and joins $COR$.

The behavior of $u$ in this case depends on whether it has an element
$(s,S)$ in $TOP.u$ such that $s=a$. Since $u \not \in COR$, if $(a,S)
\in TOP.u$, then $S$ differs from $A$. In this case if $u$ receives
$(a,A)$ then it sets $detect=\textbf{true}$. This preserves the
validity of the predicate. Alternatively, if such an entry in $TOP.u$
does not exist, then the receipt of $(a,A)$ causes $u$ to join $COR$
and forward $(a,A)$ to all its neighbors. This preserves the predicate
as well.

Thus, Predicate \ref{InvDetect} holds in the initial state of every
computation of the program and is preserved by its every action. Which
means that this predicate is an invariant of the program.  
\QED

\begin{lemma} \label{NeighborJoin}\em
If a computation of \emph{Detector} contains a state where there is a
process $u$ that belongs to $COR$ that has a non-faulty neighbor $v$
that does not, then further in the computation, either some non-faulty
process sets $detect=\textbf{\textup{true}}$ or $v$ joins $COR$.
\end{lemma}

\Proof According to Lemma \ref{ProveInvDetect}, Predicate
\ref{InvDetect} is an invariant of the program. Hence, if $u$ belongs
to $COR$ and its non-faulty neighbor $v$ does not, then channel
$Ch.u.v$ contains a message with $(a,A)$. Due to fair message receipt
assumption, $(a,A)$ is received. Observe that if $v$ is not in $COR$
and it receives $(a,A)$, then either $v$ sets $detect=\textbf{true}$
or joins $COR$.  \QED

\begin{lemma}\label{AllJoin}\em
Every computation of \emph{Detector} contains a state where either
$detect=\textbf{\textup{true}}$ in some non-faulty process or every
non-faulty process belongs to $COR$.
\end{lemma}

\Proof 
The proof is by induction on the number of non-faulty processes
in the program. As a base case, we show that $a$ itself eventually
joins $COR$. Recall, that we assume that $a$ itself is not faulty.
Observe that the program starts in a state where $start.a$ is
$\textbf{true}$. If this is so, \emph{init} is enabled.  Moreover,
\emph{init} is the only action that sets $start.a$ to
$\textbf{false}$. Thus, \emph{init} stays enabled until executed.  By
weak fairness assumption, \emph{init} is eventually executed. When
this happens, $a$ joins $COR$.

Assume that $COR$ contains $i$: $1 \leq i< n$ processes at some state
of a computation and there is a non-faulty process that does not
belong to $COR$. We assume that the connectivity of the graph exceeds
the maximum number of faulty processes. Thus, there is a non-faulty
process $u \in COR$ that has a non-faulty neighbor $v \not \in
COR$. According to Lemma \ref{NeighborJoin}, this computation contains
a state where $COR$ contains $v$.  Thus, every non-faulty process
eventually joins $COR$.  \QED

\begin{lemma}\label{NoFake}\em
If a computation of \emph{Detector} contains a state where non-faulty
process $u$ explores a fake process $v$, then this computation
contains a state where $detect=\textbf{\textup{true}}$ in some
non-faulty process.
\end{lemma}

\Proof Observe that the only source of fake process information is a
Byzantine process.  Hence, if $u$ explores a fake process $v$, then
every path to $v$ leads through a Byzantine process. Thus, in a graph
with a fake node, the maximum number of node-disjoint paths between a
real and a fake node is no more than $k$.

According to Lemma \ref{AllJoin}, eventually, either
$detect=\textbf{\textup{true}}$ at a non-faulty process or $u$
explores every non-faulty process in the system. In this case $u$
detects that all paths to the fake node $v$ lead through no more than
$k$ processes and sets $detect=\textbf{\textup{true}}$.  \QED

\begin{lemma} \label{PathNumberOK}\em
If the system does not have a faulty process, then in every
computation, for each process, the \textbf{path\_number} of the
explored subgraph $G'$ is greater than $k$.
\end{lemma}

\Proof Observe that if there are no faulty processes, only correct
topology information is circulated in the system.  Hence, for each
process $u$, $TOP.u$ contains the subgraph of the system graph $G$.
In this case, $G'.u$ is an arbitrary set of explored processes from
$G$ and the unexplored members of their neighborhoods. By the
construction of $G'.u$, every pair of unexplored processes is
connected by an edge.

\begin{figure}[ht]
\center
\epsfig{figure=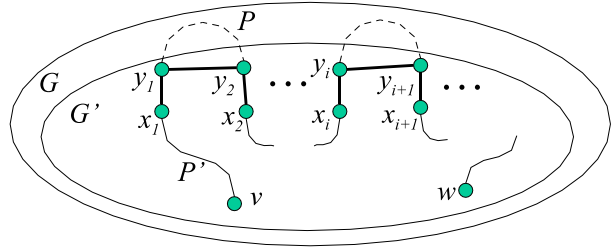,width=8cm,clip=}
\caption{Illustration for the proof of Lemma \ref{PathNumberOK}:
construction of path $P' \subset G'$ on the basis of path $P \subset
G$ }
\label{PathNumberOKFig}
\end{figure}

Let $v$ and $w$ be an arbitrary pair of explored nodes in $G'.u$.  And
let $P$ be a path connecting $v$ and $w$ in $G$. We claim that there
exists a path $P'$ in $G'.u$ connecting $v$ and $w$ that is also a
node-subset of $P$. That is, every node that belongs to $P'$ also
belongs to $P$. See Figure \ref{PathNumberOKFig} for the illustration.
If $P$ contains only the nodes explored in $G'.u$, our claim holds since
$P'=P$. Let $P$ contain unexplored nodes as well. In general, $P$
contains alternating segments of explored and unexplored nodes. Let
$\langle x_i,y_i,\cdots,y_{i+1},x_{i+1}\rangle$ be any such unexplored
segment, where $x_i,x_{i+1}$ are explored and $y_i,\cdots,y_{i+1}$ are
not. Observe that $y_i$ and $y_{i+1}$ have explored neighbors ---
$x_i$ and $x_{i+1}$ respectively. This means that both $y_i$ and
$y_{i+1}$ belong to $G'.u$. Since $y_i$ and $y_{i+1}$ are unexplored,
$G'.u$ contains an edge connecting them. We construct $P'$ to contain
every explored segment of $P$; we replace every unexplored segment by
the edge that links unexplored nodes in $G'.u$.  Observe that by
construction, $P' \in G'.u$ and $P'$ contains a subset of the nodes of
$P$. Thus, our claim holds.

Let $P_1$ and $P_2$ be two internally node disjoint paths connecting
$v$ and $w$ in $G$. According to the just proved claim, there exist
$P'_1$ and $P'_2$ belonging $G'.u$ that connect $v$ and $w$. Moreover,
$P'_1$ contains a subset of nodes of $P_1$ and $P'_2$ contains a
subset of nodes of $P_2$. Since $P_1$ and $P_2$ are internally node
disjoint, so are $P'_1$ and $P'_2$.

Recall that $G$ is assumed to be $(k+1)$-connected. This means that
for every two vertices $v$ and $w$ there exist $k+1$ internally node
disjoint paths between $v$ and $w$.  Thus, the number of internally
node disjoint paths for $v$ and $w$ in $G'.u$ is at least
$k+1$. Hence, the \textbf{path\_number} of $G'.u$ is greater than $k$.
\QED

\begin{lemma}\label{DValidity}\em 
Any computation of a detector program contains a state where a
Byzantine process is detected only if there indeed is a Byzantine
process in the system.
\end{lemma}

\Proof A non-faulty process sets $detect$ to $\textbf{true}$ if it
encounters divergent information about some node's neighborhood or
when it detects that \textbf{path\_number} is less than $k+1$.
However, a non-faulty process never modifies the neighborhood
information about other processes. Hence, if the program does not have
a faulty process, all the information about a particular neighborhood
that is circulated in the system is identical. Also, according to
Lemma \ref{PathNumberOK} if there are no faulty processes in the
system, the \textbf{path\_number} never falls below $k+1$. Hence,
$detect$ is set to $\textbf{true}$ only if indeed the system contains
a faulty process.  \QED

\begin{theorem}\em
\emph{Detector} is an adjacent-edge complete solution to the weak
topology discovery problem in case the connectivity of system topology
graph exceeds the number of faults.
\end{theorem}

\Proof To prove the theorem we show that every computation of
\emph{Detector} conforms to the properties of the problem. We then
show that the discovered topology is adjacent-edge complete.

Termination property follows from Lemma~\ref{AllJoin}, safety --- from
Lemma~\ref{NoFake}, while validity follows from Lemma~\ref{DValidity}.
Notice that Lemma~\ref{AllJoin} states that unless a fault is
detected, the neighborhood of every non-faulty process is added to
$COR$.  That is, edges adjacent to a non-faulty processes are detected
by every non-faulty processes. Thus, \emph{Detector} is adjacent-edge
complete.  Hence the theorem.\QED

\ \\ \textbf{Efficiency evaluation.} Since we consider an asynchronous
model, the number of messages a Byzantine process can send in a
computation is infinite. To evaluate the efficiency of \emph{Detector}
we assume that each process is familiar with the upper bound on the
number of processes in the system and this upper bound is in $O(n)$. A
non-faulty process then detects a fault if the number of processes it
explores exceeds this bound or if it receives more than one identical
message from the same neighbor. We assume that the process stops and
does not send or receive any more messages if it detects a fault.

In this case we can estimate the number of messages that are received
by non-faulty processes before one of them detects a fault or before
the computation terminates.  To make the estimation fair, the assume
that the unit is $log(n)$ bits. Since it takes that many bits to
assign unique process identifiers to $n$ processes, we assume that one
identifier is exactly one unit of information. A message in
\emph{Detector} carries up to $\delta+1$ identifiers, where $\delta$
is the maximum number of nodes in the neighborhood of a
process. Observe that a process can receive at most $n$ messages from
each incoming channel.  Thus, the total number of messages that can be
sent by \emph{Detector} is $2en$, where $e$ is the number of edges in
the graph. The message complexity of the program is in $O(2en\delta)$.
If $e$ is proportional to $n^2$, then the complexity of the program is
in $O(\delta n^3)$.

\section{Explorer}  
\label{SecExplorer}

\textbf{Outline.}  
The main idea of \emph{Explorer} is for each process to collect
information about some node's neighborhood such that the information
goes along more than twice as many paths as the maximum number of
Byzantine nodes. While the paths are node-disjoint, the information is
correct if it comes across the majority of the paths.  In this case
the recipient is in possession of confirmed information.  It turns out
that the topology information does not have to come directly from the
source. Instead it can come from processes with confirmed
information. The detailed description of \emph{Explorer} follows.

To simplify the presentation, we describe and prove correct the
version of \emph{Explorer} that tolerates only one Byzantine fault. We
describe how this version can be extended to tolerate multiple faults
in the end of the section.

\ \\ \textbf{Description.}  Since we first describe the 1-fault
tolerant version of \emph{Explorer} we assume that the graph is
$3$-connected.  The program is shown in Figure
\ref{FigExplorer}. Similar to \emph{Detector}, each process $p$ in
\emph{Explorer}, stores the ids of its immediate neighbors.  Process
$p$ maintains the variable \emph{start}, whose function is to guard
the execution of the action that initiates the propagation of $p$'s
own neighborhood. Unlike \emph{Detector}, however, $p$ maintains two
sets that store the topology information of the network: $uTOP$ and
$cTOP$. Set $uTOP$ stores the topology data that is unconfirmed;
$cTOP$ stores confirmed topology data.  Set $uTOP$ contains the tuples
of neighborhood information that $p$ received from other
nodes. Besides the process id and the set of its neighbor ids, each
such tuple contains a set of process identifiers, that relayed the
information.  We call it \emph{visited set}.  The tuples in $cTOP$ do
not require a visited set.

Processes exchange messages where, along with the neighbor identifiers
for a certain process, a visited set is propagated. A process contains
two actions: \emph{init} and \emph{accept}. The purpose of \emph{init}
is similar to that in the process of \emph{Detector}. Action
\emph{accept} receives the neighborhood information of some process
$r$, its neighborhood $R$ which was relayed by nodes in set $S$. The
information is received from $p$'s neighbor --- $q$.

First, \emph{accept} checks if the information about $r$ is already
confirmed. If so, the only manipulation is to record the received
information in $uTOP$. Actually, this update of $uTOP$ is not
necessary for the correct operation of the program, but it makes the
its proof of correctness easier to follow.

If the received information does not concern already confirmed
process, \emph{accept} checks if this information differs from what is
already recorded in $uTOP$ either in $r$ or in $R$. In either case the
information is broadcast to all neighbors of $p$. Before broadcasting,
$p$ appends the sender --- $q$ to the visited set $S$.

If the information about $r$ and $R$ has already been received and
recorded in $uTOP$, \emph{accept} checks if the previously recorded
information came along an internally node disjoint path. If so, the
information about $r$ is added to $cTOP$. In this case, this
information is also broadcast to all $p$'s neighbors. Note, however,
that $p$ is now sure of the information it received. Hence, the
visited set of nodes in the broadcast message is empty.

\begin{figure*}[ht]
\centering
\begin{tabbing}
1\=12345\=12345\=12345\=12345\=1234\=1234\=1234\=12345\=12345\=\kill
\> $\textbf{process}\ p$ \\
\> $\textbf{const}$ \\ 

\> \>  $P$, set of neighbor identifiers of $p$\\

\> $\textbf{parameter}$ \\
\> \>  $q: P$   \\

\> $\textbf{var}$ \\
\> \>$start:$ boolean, initially \textbf{true}, 
      controls sending of  $p$'s neighbor ids \\
\> \>$cTOP:$ set of tuples, initially $\{(p,P)\}$,  \\ 
\> \>\> (process id, neighbor id set) confirmed topology info\\
\> \>$uTOP:$ set of tuples,  initially $\varnothing$, \\
\> \>\> (process id, neighbor id set, visited id set) \\
\> \>\>  unconfirmed topology info\\
\> \>$*[$ \\
\> \emph{init}:\>\>$start \longrightarrow $ \\
\> \>\>\>$start := \textbf{false}$, \\
\> \>\>\>$(\forall j: j \in P :\textbf{send}\ (p,P,\varnothing)\ \textbf{to}\ j)$ \\
\> \>\ \ \MYBOX\\
\> \emph{accept}:\>\>$\textbf{receive}\ (r,R,S)\ \textbf{from}\ q \longrightarrow$ \\
\> \>\>\>$\textbf{if}\ (\forall t,T: (t,T) \in cTOP : t \neq r)\ \textbf{then}$ \\
\> \>\>\>\>$\textbf{if}\ (\forall t,T,U: (t,T,U) \in uTOP: t \neq r \vee T \neq R)
                       \ \textbf{then}$ \\
\> \>\>\>\>\>$(\forall j: j \in P:
                      \textbf{send}\ (r, R, S \cup \{q\})\ \textbf{to}\ j)$ \\
\> \>\>\>\>$\textbf{elsif}\ (\exists t,T,U: (t,T,U) \in uTOP: $ \\
\> \>\>\>\>\> $\;t=r \wedge R=T \wedge ((U \cap (S \cup \{q\}))) \subset \{r\})) $ \\
\> \>\>\>\>$\textbf{then}$ \\
\> \>\>\>\>\>$cTOP :=cTOP \cup \{(r,R)\}$, \\
\> \>\>\>\>\>$(\forall j: j \in P: 
                              \textbf{send}\ (r, R, \varnothing)\ \textbf{to}\ j)$ \\
\> \>\>\>$uTOP := uTOP \cup \{(r, R, S \cup \{q\})\}$ \\

\> \>\ \ $]$ \\
\end{tabbing}
\caption{Process of \emph{Explorer}} \label{FigExplorer}
\end{figure*}

\ \\\textbf{Correctness proof}. Just like for the \emph{Detector}
algorithm, we are focusing on the propagation of the neighborhood
information $A$ of a singular non-faulty process $a$. Notice that we
use $A$ to denote the correct neighborhood info. We use $A'$ for the
neighborhood information of $a$ that may not necessarily be correct.

To aid us in the argument, we introduce an auxillary set \emph{SENT}
to be maintained by each process. Since this set does not restrict the
behavior of processes, we assume that the Byzantine process maintains
this set as well. \emph{SENT} contains each message sent by the
process throughout the computation. Notice that $uTOP$ records every
message received by the process in the computation. Hence, the
comparison of $uTOP$ and $SENT$ allows us to establish the channel
contents.

Since, a message cannot be received without being sent and vice versa,
the following proposition states the invariant of the predicate that
affirms it.

\begin{proposition}\label{ProveInvExplore}\em
The following predicate is an invariant of the \emph{Explorer} program.
\begin{equation}\label{InvExplore}
\begin{split}
&(\forall b,  \textup{non-faulty}\ c, A', V : c \in B : \\
& (((a,A',V) \in Ch.b.c) \vee \\
&  ((a,A',V \cup \{b\}) \in uTOP.c)) \Leftrightarrow \\ 
& ((a,A',V) \in SENT.b))
\end{split}
\end{equation}
\end{proposition} 

The predicate states that for any process $b$ and its non-faulty
neighbor $c$ the information about the neighborhood of $a$ is recorded
in $SENT.b$ if and only if this information is en route from $b$ to
$c$ or is recorded in $uTOP.c$ with $b$ appended to the sequence of
visited nodes $V$.

Before we proceed with the correctness argument we have to introduce
additional notation.  We say that some process $c$ \emph{confirms}
$(a,A')$ if it adds this tuple to $cTOP.c$.
We view the propagation of $A'$ as construction of a \emph{tree} of
processes that relayed $A'$. This tree \emph{carries} $A'$. A tree
contains two types of nodes: a root and non-root. If process $c$ is
non-root, then for some $V$, $(a,A',V) \in SEND.c$ and $(a,A',V) \in
uTOP.c$.  That is, a non-root is a process that forwarded the
information received from elsewhere without alteration. If $c$ is a
root, then $(a,A',V) \in SEND.c$ but $(a,A',V) \not\in uTOP.c$.  Node
$c$'s \emph{ancestor} in a tree is the node that lies on a path from
$c$ to the root.

Observe that the root of a tree can only be the process $a$ itself,
the Byzantine node or a node that confirms $(a,A')$.  Notice also that
since each non-faulty process $c$ sends a message about $a$'s
information at most twice, $c$ can belong to at most two
trees. Moreover, $c$ has to be the root of one of those trees.

The proposition below follows from Proposition~\ref{ProveInvExplore}.

\begin{proposition}\label{TreeOK}\em
If some process $d$ is the ancestor of another process $c$ in a tree
carrying $(a,A')$ and $(a,A',V) \in uTOP.c$, then $d \in V$.
\end{proposition}

\begin{lemma} \label{GoodInfoInCTOP}\em
If a non-faulty node $c$ confirms $(a,A')$, then $A'=A$ and $a$ is
real.
\end{lemma}

\Proof Let us first suppose that $a$ is real.  Further, suppose $c$ is
the first non-faulty process in the system, besides $a$, to confirm
$(a,A')$.  To add $(a,A')$ to $cTOP.c$ any process $c \neq a$ has to
contain $(a, A', V) \in uTOP.c$ and receive a message from one of its
neighbors $b$ carrying $(a,A',V')$ such that $V \cap V' \subset
\{a\}$.  In our notation this means that $c$ belongs to a tree that
carries $(a,A')$ and receives a message from $b$ (possibly belonging
to a different tree) that carries the same information: $(a,A')$.  Let
us consider if $b$ and $c$ belong to the same or different trees.

Suppose $b$ and $c$ belong to the same tree. If this is the case the
messages that $c$ receives have to share nodes in the visited sets $V$
and $V'$. However, for $c$ to confirm $(a,A')$ the intersection of $V$
and $V'$ has to be a subset of $\{a\}$. That is, the only common node
between the two sets is $a$.  Observe that $a$ does not forward the
information about its own neighborhood if it receives it from
elsewhere. Thus, if $a$ belongs to a tree then $a$ is its root.  In
this case $A'=A$.

Suppose $b$ and $c$ belong to different trees. Recall that for $c$ to
confirm $(a,A')$, both of these trees have to carry $(a,A')$.
However, if $A' \neq A$ then the root of the tree is either the faulty
node or another node that confirmed $(a,A')$. Yet, we assumed that $c$
is the first node to do so. Thus, if $c$ receives a message from $b$,
the only tree that carries the information $(a,A')$ such that $A' \neq
A$ is rooted in the faulty node. Thus, even if $b$ and $c$ belong to
different trees, $A'=A$.

Similarly, if $a$ is fake, unless another node confirms $(a,A')$ there
is only one tree that carries $(a,A')$ and it is rooted in the faulty
node. In this case, no other node confirms $(a,A')$.  \QED

\begin{lemma}\label{AllJoinTree}\em
Every computation of \emph{Explorer} contains a state where each
non-faulty process belongs to at least one tree carrying $(a,A)$.
\end{lemma}

\Proof
We prove the lemma by induction on the number of nodes in the
system. To prove the base case we observe that the \emph{init} action
is enabled in $a$ in the beginning of every computation. This action
stays enabled unless executed. Thus, due to weak-fairness of action
execution assumption, \emph{init} is eventually executed in $a$. When
it is executed, $a$ forms a tree carrying $(a,A)$.

Let us assume that there are $i$: $1 \leq i <n$ non-faulty nodes that
belong to trees carrying $(a,A)$. Since the network is at least
$3$-connected, there exists a non-faulty process $c$ that
does not belong to such a tree but has a neighbor $b$ that does.

If $b$ belongs to a tree carrying $(a,A)$ then $SEND.b$ contains an
entry $(a,A,V)$ for some set of visited nodes $V$. If $c$ does not
belong to such a tree then, by definition, $(a,A,V') \not\in
uTOP.c$. In this case, according to Proposition~\ref{ProveInvExplore},
$Ch.b.c$ contains $(a,A,V)$. Similar argument applies to the other
neighbors of $c$ that belong to trees carrying $(a,A)$. That is, $c$
has incoming messages from every such neighbor.

According to the fair message receipt assumption, these messages are
eventually received. We can assume, without loss of generality, that
$c$ receives a message from $b$ first. Since $c$ does not contain an
entry $(a,A,V')$ in $uTOP.c$, upon receipt of the message from $b$,
$c$ sends a message with $(a,A,V \cup \{b\})$, attaches this message
to $SEND.c$ and includes it in $uTOP.c$. This means that $c$ joins the
tree carrying $(a,A)$.  

Thus, every non-faulty node eventually joins a tree carrying correct
neighborhood information about $a$.  \QED

A \emph{branch} of a tree is either a subtree without the root or the
root process alone. The following proposition follows from
Proposition~\ref{ProveInvExplore}.

\begin{proposition}\label{2TreesMerge}\em
If a computation of \emph{Explorer} contains a state where a
non-faulty node $c$ and its neighbor $b$ either belong to two
different trees carrying the same information $(a,A)$ or to two
different branches of the tree rooted in $a$, then this computation
also contains a state where $c$ confirms $(a,A)$.
\end{proposition}

\begin{lemma}\label{AllConfirm}\em
Every non-faulty process $c$ eventually confirms $(a,A)$.
\end{lemma}

\Proof
The proof is by induction on the number of nodes in the system.
The base case trivially holds as $a$ itself confirms $(a,A)$ in the
beginning of every computation.  Assume that $i$ non-faulty processes
have $(a,A)$ in $cTOP$, where $1 \leq i < n$. We show that if there
exists another non-faulty process $c$, it eventually confirms $(a,A)$.
Two cases have to be considered: there exists only one tree carrying
$(a,A)$, and there are multiple such trees.

Let us consider the first case. Notice, that in every computation
there eventually appears a tree rooted in $a$. In this case, we may
only consider a tree so rooted.  Since the network is at least
$3$-connected, there exists a simple cycle containing $a$ and not
containing the faulty process. According to Lemma \ref{AllJoinTree},
every process in the cycle eventually joins this tree.  Observe that,
by our definition of a tree branch, there always is a pair of neighbor
processes $b$ and $c$ that belong to different branches of a tree
rooted in $a$ and carrying $(a,A)$.  In this case, according to
Proposition~\ref{2TreesMerge}, one of the two nodes eventually
confirms $(a,A)$.

Let us now consider the case of multiple trees carrying $(a,A)$.
Again, according to Lemma \ref{AllJoinTree}, each non-faulty process
in the system joins at least one of these trees. Since the network is
at least $3$-connected there exists a non-faulty process $c$ belonging
to one tree that has a neighbor $b$ belonging to a different tree. In
this case, according to Proposition~\ref{2TreesMerge}, $c$ confirms
$(a,A)$.

By induction, every non-faulty process in the system eventually
confirms $(a,A)$.  \QED

\begin{theorem}\em
\emph{Explorer} is a two-adjacent-edge complete solution to the strong
topology discovery problem in case of one fault and the system
topology graph is at least $3$-connected.
\end{theorem}

\Proof \emph{Explorer} conforms to the termination and safety
properties of the problem as a consequence of Lemmas~\ref{AllConfirm}
and \ref{GoodInfoInCTOP} respectively.

Observe that a non-faulty node may potentially confirm incorrect
neighborhood information about a Byzantine node. That is, an edge
reported by the faulty process is either missing or fake. However, due
to the two above lemmas, if two nodes are non-faulty the information
whether there is an adjacent edge between them is discovered by every
non-faulty node. Hence \emph{Explorer} is two-adjacent-edge complete.
\QED

\ \\ \textbf{Modification to handle $k>1$ faults.}  Observe that
\emph{Explorer} confirms the topology information about a node's
neighborhood, when it receives two messages carrying it over
internally node disjoint paths. Thus, the program can handle a single
Byzantine fault.  \emph{Explorer} can handle $k>1$ faults, if it waits
until it receives $k+1$ messages before it confirms the topology
info. All the messages have to travel along internally node disjoint
paths. For the correctness of the algorithm, the topology graph has to
be $(2k+1)$-connected.

\begin{proposition}\em
\emph{Explorer} is a two-adjacent-edge complete solution to the strong
topology discovery problem in case of $k$ faults and the system
topology graph is at least $(2k+1)$-connected.
\end{proposition}

\noindent
\textbf{Efficiency evaluation}. Unlike \emph{Detector},
\emph{Explorer} does not quit when a fault is discovered. Thus, the
number of messages a faulty node may send is arbitrary large. However,
we can estimate the message complexity of \emph{Explorer} in the
absence of faults. Each message carries a process identifier, a
neighborhood of this process and a visited set. The number of the
identifiers in a neighborhood is no more than $\delta$, and the number
of identifiers in the visited set can be as large as $n$. Hence the
message size is bounded by $\delta+n+1$ which is in $O(n)$.

Notice, that for the neighborhood $A$ of each process $a$, every
process broadcasts a message twice: when it first receives the
information, and when it confirms it. Thus, the total number of sent
messages is $4e \cdot n$ and the overall message complexity of
\emph{Explorer} if no faults are detected is in $O(n^4)$.

\section{Composition and Extensions}
\label{SecExtensions}

\textbf{Composing \emph{Detector} and \emph{Explorer}}.  Observe that
\emph{Detector} has better message complexity than \emph{Explorer} if
the neighborhood size is bounded. Hence, if the incidence of faults is
low, it is advantageous to run \emph{Detector} and invoke
\emph{Explorer} only if a fault is detected. We assume that the
processes can distinguish between message types of \emph{Explorer} and
\emph{Detector}.  In the combined program, a process running
\emph{Detector} switches to \emph{Explorer} if it discovers a
fault. Other processes follow suit, when they receive their first
\emph{Explorer} messages.  They ignore \emph{Detector} messages
henceforth.  A Byzantine process may potentially send an
\emph{Explorer} message as well, which leads to the whole system
switching to \emph{Explorer}. Observe that if there are no faults, the
system will not invoke \emph{Explorer}. Thus, the complexity of the
combined program in the absence of faults is the same as that of
\emph{Detector}. Notice that even though \emph{Detector} alone only
needs $(k+1)$-connectivity of the system topology, the combined
program requires $(2k+1)$-connectivity.

\ \\ \textbf{Message Termination.}  We have shown that \emph{Detector}
and \emph{Explorer} comply with the functional termination properties
of the topology discovery problem. That is, all processes eventually
discover topology. However, the performance aspect of termination,
viz. message termination, is also of interest. Usually an algorithm is
said to be message terminating if all its computations contain a
finite number of sent messages \cite{DS80}.

However, a Byzantine process may send messages indefinitely. To
capture this, we weaken the definition of message termination.  We
consider a Byzantine-tolerant program \emph{message terminating} if
the system eventually arrives at a state where: (a) all channels are
empty except for the outgoing channels of a faulty process; (b) all
actions in non-faulty processes are disabled except for possibly the
receive-actions of the incoming channels from Byzantine processes,
these receive-actions do not update the variables of the process. That
is, in a terminating program, each non-faulty process starts to
eventually discard messages it receives from its Byzantine neighbors.

Making \emph{Detector} terminating is fairly straightforward. As one
process detects a fault, the process floods the announcement
throughout the system. Since the topology graph for \emph{Detector} is
assumed $(k+1)$-connected, every process receives such announcement.
As the process learns of the detection, it stops processing or
forwarding of the messages.  Notice that the initiation of the flood
by a Byzantine node itself, only accelerates the termination of
\emph{Detector} as the other processes quickly learn of the faulty
node's existence.

The addition of termination to \emph{Explorer} is more involved. To
ensure termination, restrictions have to be placed on message
processing and forwarding.  However, the restrictions should be
delicate as they may compromise the liveness properties of the
program.  By the design of \emph{Explorer}, each process may send at
most one message about its own neighborhood to its neighbors. Hence,
the subsequent messages can be ignored. However, a faulty process may
send messages about neighborhoods of other processes.  These processes
may be real or fake. We discuss these cases separately.

Note that each process in \emph{Explorer} can eventually obtain an
estimate of the identities of the processes in the system and
disregard fake process information.  Indeed, a path to a fake node can
only lead through faulty processes. Thus, if a process discovers that
there may be at most $k$ internally node disjoint paths between itself
and a certain node, this node is fake.  Therefore, the process may
cease to process messages about the fake node's neighborhood. Notice,
that since the system is $(2k+1)$-connected, messages about real nodes
will always be processed. Therefore, the liveness properties of
\emph{Explorer} are not affected.

As to the real processes, they can be either Byzantine or non-faulty.
Recall that each non-faulty process of \emph{Explorer} eventually
confirms neighborhoods of all other non-faulty processes. After the
neighborhood of a process is confirmed, further messages about it are
ignored.

The last case is a Byzantine process $u$ sending a message to its
correct neighbor $v$ about the neighborhood of another Byzantine
process $w$. By the design of \emph{Explorer}, $v$ relays the message
about $w$ provided that the neighborhood information about $w$ differs
from what previously received about $w$. As we discussed above,
eventually $v$ estimates the identities of all real processes in the
system. Therefore, there is a finite number of possible different
neighborhoods of $w$ that $u$ can create. Hence, eventually they will
be exhausted, and $v$ starts ignoring further messages form $u$ about
$w$.

Thus, \emph{Explorer} can be made terminating as well.

\ \\ \textbf{Handling topology updates}. In the topology discovery
problem statement, it is assumed that the system topology does not
change. However, \emph{Detector} and \emph{Explorer} can be adapted to
manage topology changes as well. There are two aspects of topology
change: the \emph{notification} and the \emph{transport}. For
notification, a node should inform the others of its most up-to-date
neighborhood. The transport aspect should ensure that this notification
is delivered to all nodes despite of topology changes.

We implement the transport aspect as follows. If a node $p$, due to the
change in topology, obtains a new neighbor $q$. Then $p$ sends to $q$
the most recent neighborhood information about all nodes that $p$ is
aware of. Thus, the most recent information gets propagated regardless
of topology changes.

The satisfaction of the notification aspect is more involved.
Observe, however, that apart from detecting fake nodes in
\emph{Explorer}, both algorithms propagate the information of one
process neighborhood independently of the others. We first describe
how this propagation can be done in case the topology changes and then
address the fake node detection. Each time the neighborhood of a
process $p$ changes, $p$ starts a new \emph{version} of the topology
discovery algorithm for its neighborhood. Observe that a faulty
process may also start a new version for $p$.

The versions are distinguished by version numbers. Each process
maintains the version numbers of $p$. Each related message carries the
version number.  Each process outputs the discovered neighborhood of
$p$ with the highest received version number. Observe that in the case
of \emph{Explorer} the processes only output confirmed information.
Notice that if a faulty process sends incorrect information about
$p$'s neighborhood with a certain version number, this incorrect info
will be handled by the basic \emph{Detector} or \emph{Explorer} within
that version. For example, the faulty messages of version $i$ about
$p$'s neighborhood will be countered by the correct messages of the
same version. Notice that a faulty process in \emph{Explorer} may
start a version $j$ for $p$'s neighborhood such that it is higher than
the highest version $i$ that $p$ itself started. However, according to
the basic \emph{Explorer}, the incorrect information in version $j$
will not be confirmed.

There are two specific modifications to the basic \emph{Detector}. If
the faulty process sends a message concerning $p$ with the version
number higher than that of $p$, $p$ itself detects the fault. To
detect fake nodes generated by a faulty process, each node has to
compile the topology $TOP$ graph of the highest version number for
each node in the system and ensure that its connectivity does not fall
below $k+1$.  Observe that \emph{Detector} is unable to differentiate
between temporary lack of connectivity from malicious behavior of the
faulty nodes. Therefore, the connectivity of the discovered network at
each node should never fall below $k+1$. For that, we assume that
throughout a computation the intersection of all system topologies is
$k+1$-connected. This assumption is not necessary for \emph{Explorer}.

The notification mechanism can be optimized in obvious ways. For
\emph{Detector}, each process has to keep the information for $p$ with
only the highest version number. Obsolete information can be safely
discarded. For \emph{Detector}, the process may keep the latest
version of \emph{confirmed} neighborhood information.  Observe that
this extension of the topology discovery algorithms assumes
infinite-size counters. Care must be taken when implementing these
counters in the actual hardware, as the faulty processes may try to
compromise topology discovery if the counter values are reused.
Hence, such an implementation would require a Byzantine-robust counter
synchronization algorithm. Lamport and Melliar-Smith \cite{LM86}
proposed such algorithm for completely connected systems. Extending it
to arbitrary topology systems is an attractive avenue of future
research.

\ \\ \textbf{Discovering neighbors.}  As described, in the initial
state of \emph{Detector} and \emph{Explorer}, each process has access
to correct information about its immediate neighborhood. Note that, in
general, obtaining this information in the presence of Byzantine
processes may be difficult as they can mount a Sybil attack
\cite{D02}. In such an attack, a faulty process is able to send a
message and put an arbitrary process identifier as the sender of this
message. That is, a faulty process \emph{assumes} the identity of this
process.  Sybil attack is difficult to handle.  However,
\emph{Detector} and \emph{Explorer} can be modified to handle
neighborhood discovery with known ports. That is, each process does
not know the identities of its neighbors but can determine if a
message is coming from the same process.

The modified algorithms contain two phases: neighborhood discovery
phase and topology discovery proper phase. In the first phase, each
process broadcasts its identifier to its neighbors.  Observe that
faulty processes may not send these initial messages at all. Thus, the
process should not wait for a message from every possible
neighbor. Instead, as soon as each process $p$ gets a message with $q$
in its identifier, $p$ may start the second phase with $\{q\}$ as its
neighborhood. Every time $p$ gets a new distinct identity, $p$ treats
it as topology update, increments its counter and re-initiates the
topology discovery. This procedure can be further streamlined. Recall
that for \emph{Detector} and \emph{Explorer} the topology graph has to
be respectively $k+1$ and $2k+1$-connected. Thus, depending on the
algorithm, each process is guaranteed to have $k+1$ or $2k+1$
non-faulty neighbors. Therefore, each process may delay initiating
topology discovery until it gets this minimum number of distinct
identities.

\begin{figure}
\center
\epsfig{figure=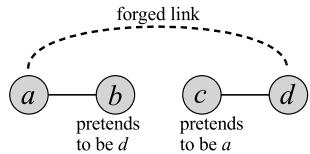,width=4cm,clip=}
\caption{Faulty nodes $b$ and $c$ forge a link between non-faulty
 nodes $a$ and $d$.}\label{blackhole}
\end{figure}

Observe that due to known ports a faulty process may not be able to
use more than one identifier per neighbor without being detected.
However, the modified algorithms may not be able to determine the
identifier of a faulty process as it may select an arbitrary one,
including the identifier of an already existing process. Thus, a pair
of colluding faulty nodes may deceive their non-faulty neighbors into
believing that they share an edge. This behavior is illustrated in
Figure \ref{blackhole}. When communicating to a non-faulty node $a$,
its faulty neighbor $b$ assumes the identity of another non-faulty
node $d$.  Similarly, a faulty neighbor $c$ of $d$ assumes the
identity of $a$. This way, non-faulty nodes $a$ and $d$ are led to
believe they share an edge.

\ \\ \textbf{Other extensions}.  Observe that \emph{Explorer} is
designed to disseminate the information about the complete topology to
all processes in the system. However, it may be desirable to just
establish the routes from all processes in the system to one or a
fixed number of distinguished ones.  To accomplish this
\emph{Explorer} needs to be modified as follows. No neighborhood
information is propagated. Instead of the visited set, each message
carries the propagation path of the message. That is, the order of the
relays is significant.

Only the distinguished processes initiate the message propagation. The
other processes only relay the messages.  Just as in the original
\emph{Explorer}, a process confirms a path to another process only if
it receives $2k+1$ internally process disjoint paths from the source or
from other confirming processes.  Again, like in \emph{Explorer}, such
process rebroadcasts the message, but empties the propagation path. In
the outcome of this program, for every distinguished process, each
non-faulty process will contain paths to at least $2k+1$ processes
that lead to this distinguished process.  Out of these paths, at least
$k+1$ ultimately lead to the distinguished process.

 In \emph{Explorer}, for each process the propagation of its
neighborhood information is independent of the other neighborhoods.
Thus, instead of topology, \emph{Explorer} can be used for efficient
fault-tolerant propagation of arbitrary information from the processes
to the rest of the network.

\section{Conclusion}\label{SecEnd}

In conclusion, we would like to outline a couple of interesting
research directions.  The existence of Byzantine-robust topology
discovery solutions opens the question of theoretical limits of
efficiency of such programs.  The obvious lower bound on message
complexity can be derived as follows. Every process must transmit its
neighborhood to the rest of the nodes in the system.  Transmitting
information to every node requires at least $n$ messages, so the
overall message complexity is at least $\delta n^2$.  If $k$ processes
are Byzantine, they may not relay the messages of other nodes. Thus,
to ensure that other nodes learn about its neighborhood, each process
has to send at least $k+1$ messages.  Thus, the complexity of any
Byzantine-robust solution to the topology discovery problem is at
least in $\Omega(\delta n^2 k)$.

Observe that \emph{Explorer} and \emph{Detector} may not explicitly
identify faulty nodes or the inconsistent view of the their immediate
neighborhoods. We believe that this identification can be accomplished
using the technique used by Dolev \cite{D82}.  In case there are
$3k+1$ non-faulty processes, they may exchange the topologies they
collected to discover the inconsistencies. This approach, may
potentially expedite termination of \emph{Explorer} at the expense of
greater message complexity: if a certain Byzantine node is discovered,
the other processes may ignore its further messages.

\bibliographystyle{plain}
\bibliography{topology}

\end{document}